\documentclass[12pt]{article}

\begin{document}

 Bose Fermi Supersymmetry with Bogoliubov transforms in Cosmology

\qquad\qquad\qquad\qquad \qquad Ajay Patwardhan

\qquad\qquad\qquad Physics Department, St Xaviers college,

\qquad\qquad\qquad Mahapalika Marg, Mumbai, India

\qquad\qquad\qquad\qquad\qquad ABSTRACT

Field theory including Supersymmetry and Bose Fermi symmetry is an active subject of particle physics and cosmology. Recent and expected  observational evidence gives indicators for the creation and destruction of normal and supersymmetric dark matter in the universe. This paper uses Bogoliubov transforms in supersymmetric and Bose Fermi form for obtaining the vaccuum expectation values at any  two times in cosmological and black hole  geometries. The isotropic Robertson Walker and slightly anisotropic Bianchi I geometry mode functions have a differential equation form analogous to the supersymmetric hamiltonian. The conditions for mixed and distinct representations for bosonic and fermionic fields of normal and supersymmetric partner particles are found.

\qquad\qquad\qquad  \qquad 1.INTRODUCTION

One of the fundamental symmetries of nature is expected to be Supersymmetry(SUSY). The dark matter in the universe could be neutralinos which are likey to be discovered in LHC experiments in a few years. Supersymmetric partners and the normal particles have asymmetric production and decay , possibly from the inflation and post inflation epoch. The primordial black holes and the collapsed matter black holes in both dark matter and normal matter are active subjects of research. The Fermi Bose symmetry , or exchange symmetry is a subset of the SUSY and also occurs in nuclear physics and in condensed matter physics. This occurs in the form of an exchange Hamiltonian giving rise to hyperfine structure in nuclear spectra and as  Fermi Bose joint quantisation in anyonic systems in condensed matter. The quantum group representations give a commutator and anti commutator algebra ; a $``q''$ commutation  algebra; of which the Fermi and Bose algebras are special cases. Grassmanian and normal co-ordinates can be used to express the Fermionic and Bosonic systems in superspace and do SUSY Yang Mills gauge theories, while  superstring  and supergravity theories have SUSY built in.[Ref 1 to 22]

 Local Lorentz invariance, asymptotically free field theories and the CPT theorem and possibly locality are usually assumed to hold. This gives the normal particles/SUSY particles a definite spin, half integer or integer, and spin statistics theorems prohibit mixing of the spin multiplets. However in rapidly accelerating frames such as during inflationary cosmology and in collapsing stars into black hole  horizons at the singularity ; and in interacting field theories it is possible that the conditions to mix the spin multiplets can arise. This is expressed as generalised Bogoliubov transforms on the commutation algebra. In the particular geometries in the early universe and in late collapse of black holes, these transforms can be calculated and the conditions derived that mix or do not mix the spin multiplets of Fermions and Bosons. The Fermi Bose symmetry breaking gives an insight into the interior of black holes as well as in the creation of normal and dark matter in cosmology.[Ref 1 to 22]

In this paper the framework of supersymmetry and distinct and mixed Bose Fermi representations are created in curved spacetime using Robertson Walker and Bianchi I geometries. The Bogoliubov transforms between any two times for the commutation preserving  algebra of operators are obtained. This gives conditions on the creation and destruction of normal and supersymmetric partner particles of bosonic, fermionic type. The supersymmetric hamiltonian is seen to have a similar structure to the evolution equation for the field modes in these geometries. This allows a construction of a specific framework for quantum fields in the universe.

\qquad 2 SUPER SYMMETRIC FUNDAMENTAL FIELD THEORY

In SUSY Yang Mills field  theories the consistency of symmetries and structures arising from them was developed in  earlier work [Ref 2 to 5]. The combined representation of fermionic (finite sector) and bosonic (infinite sector) require a  Grassmannian and normal space. This occurs naturally in quantum groups on the operator algebra.

 The mixed systems with $b^{\dagger}_{\alpha} $,\qquad $\alpha = 1,2,...n$ for Bose case

 and the $f^{\dagger}_{i}$,\quad $ i=1,2...m$ for Fermi case give rise to a

 supersymmetric transform in $b^{\dagger}b, f^{\dagger}f$ in Bose sector and $f^{\dagger} b, b^{\dagger} f $ in Fermi sector expressed as a matrix operator

\begin{displaymath}
\left(\begin{array}{c|c}
b^{\dagger}b & b^{\dagger} f\\
f^{\dagger}b & f^{\dagger}f
\end{array}\right)
\end{displaymath}

The matrix Bogoliubov transforms act on this operator causing mixed sectors.

Exchange terms in Hamiltonians for interacting Bose and Fermi systems in nuclei have such terms creating hyperfine spectral lines. [Ref 12]. 

In the early universe it is expected that broken SUSY allows the Fermi and Bose sectors to decouple and superselection rules apply that prohibit mixed state representations due to local Lorentz invariance. However in the accelerated geometries we have the possibility of violating these conditions and permit more general Bogoliubov transforms on each sector as well as on the whole system.

In the $N$ particle Hilbert space $H_{N}$ for indistinguishable particle states the permutation group $ S _{N} $ acts with a [Ref 2,3]

 projection operator defined by $P_{\theta} = \frac{d(\theta)}{N!}\sum_{(s \in S_{N})} \chi_{\theta}^{c_{js}} (s) $

 where character of $\theta$ and dimension of $\theta$ are inserted.

 Then $ H_{N} (\theta) = P_{\theta} H_{N} $ 

and a particular choice of the irreducible representation $\theta$ of $ S_{N} $ is made.

 The superselection sectors correspond to various $ \theta$. The union over $\theta$ of the projection operators is the total set of observables. These generate a $\theta$ statistics for a fixed $\theta$.

 The Bose and Fermi case are the special cases of representations of $S_{N} $  with $\frac{1}{N!} \sum s$ and $\frac{1}{N!} \sum sign(s)$.

 The $ Z_{2} $ graded $H_{N} $ then gives the standard symmetric and antisymmetric case for quantum states. The generalised Bogoliubov transforms extend to these observable algebras in $ d $ and $ N-d$ dimensions.

$U^{(b)}_{(d)}\otimes U^{(f)}_{(N-d)}$ is the direct product representation. With SUSY it is possible to obtain a doubling of particle states as normal and SUSY partners and that allows a swap of the statistics.

 For the  Fermi to Bose and Bose to Fermi exchange of  statistics transforms, superspace coordinates are convenient to give a unified framework. The `` No superposition between states belonging to different sectors `` condition can be relaxed in interacting field theories with no asymptotic free particle states, in curved space time  geometry with finite time Bogoliubov transforms and for any non inertial frames with no unique vaccuum state.

$ Q\mid B> =\mid F>$ and $ Q^{\dagger}\mid F> = \mid B>$ operators are required.

3. THE BOGOLIUBOV TRANSFORMS ON THE FERMI BOSE SYSTEMS

The creation and anhilation operators $f^{\dagger}$ and $f$ transform as

$ \bar {f} = \alpha f +\beta   f^{\dagger}$ ; 

  and similarly the  adjoint case, and for the bosonic $b$ operators . 

This transform is anti  commutator  and commutator 
algebra preserving for the conditions

$\mid \alpha \mid^{2}  \pm \mid \beta \mid ^{2} = 1 $

for Fermi$ (+)$ and Bose$ (-)$  sign cases

The number operator is $b^{\dagger}b$ for bosonic and  $f^{\dagger} f $ for fermionic case and

 its expectation value is 

$<0\mid N \mid 0> = = \mid \beta \mid^{2}$

 More generally the $\alpha $ and $ \beta$ can be matrices for multimode systems. The transforms with the $q$ deformed commutation were also given in [Ref 17].

The traceclass and Hilbert Schmidtt  condition on the $\beta \beta^{\dagger}$ is required. 

A more general Bogoliubov transform can be defined that acts on the supersymmetric matrix operators above and can give distinct as well as mixed sectors as special cases.

\begin{displaymath}
\left(\begin{array}{c|c}
\phi & \psi \\
\bar \psi & \bar\phi
\end{array}\right)
\end{displaymath}

and its inverse can be defined to make the Bogoliubov transforms invertible. 
Ref(Berezin)

$ \psi  \psi^{\star} $ traceclass, Hilbert Schmidtt

 and $\phi  \phi^{\star} $ has a Fredholm determinant 

give a proper canonical transform iff 

$ \phi \phi^{\star} = I + \psi  \psi^{\star} $

For matrix operators $\phi$ and $\psi$ this property becomes

$\phi\phi^{\dagger} = I + \psi \psi^{\dagger}$ 

and $\phi \psi^{tr} = \psi \phi^{tr} $

With these the system of many bosonic and fermionic variables can be treated.

Using the analysis of bosonic and fermionic coherent and squeezed states generated by Unitary operators involving products of creation and anhillation operators, denoted $ B $, it is possible to write this general Bogoliubov transform of matrix form as a unitary operator with a angle $[0,\pi]$ such that $Exp (iB)$ has fermionic or bosonic or mixed sector representations. The Baker Campbell identity for exponentiated products of operators has to be used.

 Positive energy solutions for Fermions and vanishing commutators at spacelike separations for Bosons, with a vaccuum state defined for each allowed and transformed $ f\mid0> =  0 $,$b\mid0> = 0$

Mixed operators can be defined as $fb^{\dagger} \pm b^{\dagger}f $ 

and $f^{\dagger}b \pm bf^{\dagger}$

for the bosonic $b$ and fermionic $ f$ operators and their adjoints.

Also $ [f,f]_{\pm}  $, $[f,b]_{\pm} $ , $[b,b]_{\pm} $, and their adjoints;

and similarly for the operators' algebra with their adjoints  is defined such that

 $b^{\dagger}f$, $f^{\dagger}b $ commute with the $H$

 such that $[Q,H]_{-} = 0 $ for the supersymmetric charge operator $Q$.

This $ Q=bf^{\dagger} $ and $ Q^{\dagger} = b^{\dagger}f $

$[Q,Q^{\dagger}]_{+} = H $; and$ [Q,H]_{-} = 0 $,$[Q^{\dagger},H]_{-} = 0 $

while $[Q,Q]_{+} = 0 = [Q^{\dagger},Q^{\dagger}]_{+}$
 
This generates a closed system of supersymmetric algebra with the supersymmetric hamiltonian $H$ and charge $Q$.

The general Bogoliubov transforms act upon these operators as automorphisms preserving the algebraic structure.

 The half integer fermionic and integer bosonic representations have  equal time commutation relations   as propagators in field theory. 

These  get modified for two time commutations between two spacelike hypersurfaces in geometries that have timelike Killing vector, every where orthogonal to spacelike hypersurfaces, using Bogoliubov two time transforms. A particular supersymmetric representation of these operator algebras and their Bogoliubov transforms has a correspondence with the equation for the field modes in the curved spacetime geometries as shown in this paper. Hence there is a physically significant realisation of the mathematical structures in our universe.

4. RIEMANNIAN SPACE TIME OF GENERAL RELATIVITY AND BOSE FERMI COMMUTATION ALGEBRA

 Fulling and Birrell and Davies [Ref 1,6] gave conditions on Bogoliubov transforms in curved space time. For transforms taken at two times in the evolving universe a cosmological model scale factor , such as in Robertson Walker (isotropic) or Bianchi I (anisotropic) model could be used. These  metrics  are also useful in the interior solution of a black hole for end stage of collapse.

The Number operator is found in vaccuum states at two times. This is a situation for the beginning time from big bang  to a later time ; or black hole formation to evaporation, and between  any two intermediate times in universe, and from any time to the end stage of collapse time in black holes.

The accelerated geometry in all these cases gives non zero values for the number operator , contributing to particle creation. 

Bosonic operator algebra is commutative and related to symplectic and Lorentz groups while Fermionic operators are anticommutative related to rotational and unitary symmetry. The vaccuum expectation values are respectively antisymmetric and symmetric in their arguments. These differences are at the level of the operators and their representations. The automorphisms that leave these structures invariant are the Bogoliubov or linear canonical transforms. The Hilbert space of states have the permutation group projection operators and superselection rules on the Bosonic and Fermionic sectors in free field theories in flat space time. These can be generalised to mixed, and  direct product representations in Supersymmetry for interacting fields in curved spacetime as shown below.

$<0^{(t_{1})}\mid N^{(t_{2})} \mid 0^{(t_{1})}> = \beta \beta^{\dagger} $

 where $\beta$ is a function of the accelerated geometry scale factor and the two times $ t _{1},  t_{2}$

This can be written as $<0^{t_{2}} \mid T_{00}(t_{2} , x) \mid 0^{t_{2}}> = 0$ and 

$ <0^{t_{1}}\mid T_{00}(t_{2} , x)\mid 0^{t_{1}}> \neq 0 $

for the stress energy tensor.

In supersymmetric quantum theory we define the operator $ Q $ such that 

$[Q,Q^{\dagger} ]_{+} = QQ^{\dagger} + Q^{\dagger}Q = H$ 

We could write all possible bilinear quadratic products of the bosonic and fermionic operators and their adjoints. 

This constitutes the complete set of commutation sub algebra.

 For example $f^{\dagger} b $,\qquad $b^{\dagger} f$,\qquad$ fb^{\dagger} $,\qquad$ bf{\dagger}$ 

and various hermitian and anti hermitian linear combinations.
A hamiltonian can be defined for the supersymmetric charge $Q$ and its representation found in terms of the bosonic and fermionic operators.

with $ [H,Q]_{-} = 0 =[H,Q^{\dagger} ]_{-}$

The hamiltonian $H$ can have a representation as 

$ H = ( -\frac{d^{2}}{dx^{2}} + W^{2}(x)) I - \sigma_{z} \frac{dW}{dx} $

As seen in the next section in a number of cases of curved space geometry,  the differential equation obtained for the mode functions $\chi_{k} (\eta) $ are of this form. Hence they are like  eigenfunctions of the supersymmetric hamiltonian.

 Hence the representation for the Bose , Fermi operators can be obtained as 

Bosonic : $ b = \frac{d}{dx} + W(x) $ and the adjoint $ b^{\dagger} = \frac{-d}{dx} + W(x) $ 

and the fermionic: $ f = \sigma_{+} $ and $ f^{\dagger} = \sigma_{-} $

 with $\sigma_{(z,\pm)}$  being the Pauli spin matrices of $SU(2)$

The commutator of the $ b $ and $b^{\dagger} $ equals to $\frac{dW}{dx} $ 

and anticommutator of the $f$ and $f^{\dagger} $ equals to $1$.

 Then the $ Q= bf^{\dagger} $ and $ Q^{\dagger} = b^{\dagger} f = -ibf^{\dagger} $ 

for $[f,f^{\dagger}]_{-} =\sigma_{z} $

These relations can be extended for multi mode systems as matrices using direct products between the operators; and they represent bosonic anhilation with fermionic creation for $Q$ and the converse for $Q^{\dagger} $, as required .  Mixed state representations can be obtained by using the generalised Bogoliubov transform.

 The formal similarity  between the supersymmetric hamiltonian and the evolution equation for the mode functions $\chi$ as seen in next section , with the variable $x$ replaced by $\eta$ allows construction of supersymmetric Bogoliubov transforms in curved spacetime. Moreover  some of the neccessary conditions for non violation of superselection rules become relaxed or invalid in curved spacetime. 

With accelerated geometry and non inertial reference frames, in which to define quantum states, we can have the possibility of supersymmetric sector mixing Bogoliubov transforms. Hence the mechanism for creation of both normal particles and supersymmetric partners of particles exists.

5. SUSY BOGOLIUBOV TRANSFORMS ON CURVED SPACETIME

The formal similarity of the supersymmetric hamiltonian with the evolution equations for modes of the fields in the cosmological spacetime geometry is intriguing. As initially the supersymmetry generators were defined as an extended Poincare group algebra , that is a spacetime related property. Later in Yang Mills gauge theory the SUSY was related to the internal symmetries, with graded Lie algebras and superspace. However taking the view that a 10 dimensional space with six dimensional compactified Tian Yau or Calabi Yau spaces with a direct product with a four dimensional Riemannian spacetime is the theory of the universe ; consistent with observations in high energy physics and astrophysics, it appears that SUSY provides the link between internal(dynamical) and kinematical (external) symmetries in a unified fundamental field theory in ten dimensions.[Ref 17] 

In the Robertson Walker and Bianchi I geometries the mode functions $\chi_{k} (\eta)$ transform and give the Bogoliubov transform coefficients

The $\beta $ factors  in the Bogoliubov transforms give the condition on number operators to have expectation values given by

$\mid\beta\mid^2 = \frac {Sinh^{2}(\frac{(\pi \omega_{-})}{\rho})}{Sinh(\frac{\pi \omega_{i}}{ \rho}) Sinh(\frac{\pi \omega_{f}}{ \rho})} $

where  $ \omega_{\pm} = 0.5(\omega_{i} \pm \omega_{f} ) $ ;

and $\omega_{i} $ is  frequency at initial time and $\omega_{f} $ is  frequency at final time.

The scale factor$ a(t) $ of the geometry $ dS^{2} = dt^{2} - a^{2} (t) dx^{2}$

 with $\eta = \int\frac{1}{a(t)} dt$

and the $C(\eta ) = a^{2} (\eta)$ with

 $ \int a(\eta) d \eta = \int^{t} dt$ as the scale factor

 and $C(\eta) = a^{2}(t).$

 Then $ c(\eta) = A + B  tanh (\rho \eta) $ becomes $ A \pm B $ as $ \eta $ becomes $\pm $ infinity. 

Introducing three parameters $ A, B, \rho $ we get the mode functions that satisfy the equations

 $\frac{d^{2} \chi_k} {d \eta^{2} } + (k^{2} +  C(\eta) m^{2} ) \chi_k (\eta) = 0$

giving  $\omega_{i} = ( k^{2} +m^{2} (A -B))^{1/2} $ 

and $\omega_{f} = ( K^{2} + m ^{2} ( A +B ) )^{1/2}$

It is these equations that can be compared with the supersymmetric Hamiltonian , with the $W^{2}(\eta)$ and $\frac{dW(\eta)}{d\eta} $ terms. In that notation the collection of all operators $b , f, Q, H $ and their adjoints were constructed and now these have a specific realisation in terms of the field mode function equation in the curved space time geometry.

 These mode functions allow construction of the coefficients for the Bogoliubov transforms on the operators. 

For example, $b(t_{f}) = \alpha_{(t_{f},t_{i})}  b(t_{i}) +  \beta_{(t_{f},t_{i})}^{\star} b^{\dagger}(t_{i})$

Then the $\beta \beta^{\star} (t_{f},t_{i} )$ gives the non zero expectation values and particle creation.

Following [REF 1] this gives the required $\mid \beta \mid^{2} $ as above. 

The particle creation rate can be found. Depending on the Fermi or Bose fields used these expressions are found for normal and supersymmetric partners.

In the cosmological models with a $10^{-5}$ anisotropy as seen from WMAP data, the Bianchi I model with slight anisotrpic case gives the same expression as above with the $m^{2} $ term modified by

 $ \alpha^{2} cosec^{2} (\eta) $ where $\alpha$ is the anisotropy parameter.

 A similar expression will be used for slightly anisotropic collapse in interior solution of a black hole geometry.

Similar to the number operator  expectation values , we can also find the  vaccuum expectation values of components of the stress energy tensor

$\frac{ <0(f)\mid T_{\mu \nu} \mid  0(i)> }{<0(f) \mid 0(i)>}$ 

between any initial and final time vaccuum states in the geometry. 

The geodesically incomplete spacetimes ; that is the initial big bang and the final collapse to a singularity in black holes require that these expectation values will be non zero. We can find them for the bosonic and fermionic cases as well as the supersymmetric and mixed representation cases.

6. IMPLICATIONS FOR COSMOLOGY MODELS

In cosmology the  analysis above can be used in four cases:

1. From initial big bang to some later time

2. From collapse of primordial black hole to its evaporation

3. From any finite time to a later finite time 

4. In collapse of material in a future geodesically incomplete region.

The time dependent Bogoliubov transforms in the accelerated geometry are used in the first case, for example in the inflation with scalar field and baryogenesis epoch. The Robertson Walker or Bianchi I models are used in third case for isotropic and slightly anisotropic expansion of the universe. The interior solutions of the primordial black holes are modeled with the Robertson Walker/ Bianchi I metrics for second case. And with the initial and final time interchanged for the fourth case. The generalised supersymmetric Bogoliubov transforms are used for normal and dark matter creation and anhillation , with particles and their supersymmetric partners.[Ref 17 to 22]

7. IMPLICATIONS FOR BLACK HOLE MODELS OF COLLAPSE AND EVAPORATION

These can arise from above analysis in four cases:

1. Primordial black hole formation and evaporation

2. Collapse of stellar mass or higher mass into horizon and infall in to possible singularity

3. Black hole evaporation

4. SUSY, normal and dark matter black holes

As above the primordial black hole case is common and gives a particle and supersymmetric particle  production rate. The second case requires interior solution which is modeled as Robertson Walker for symmetric and Bianchi I for asymmetric collapse. The third case is the standard Hawking effect. In the fourth case the generalised Bogoliubov transforms are used for normal and dark bosonic and fermionic matter to be created and destroyed.[Ref 17 to 22]

\qquad\qquad \qquad 8. CONCLUSION

There are a large and growing number of publications on the astrophysical implications of supersymmetric dark matter in cosmology and black holes. This paper attempts to find a Bogoliubov transform approach to obtaining the conditions for supersymmetry and Bose Fermi symmetry in curved spacetime. This will enable the understanding of the collapse to a black hole in which Fermion and Boson number are not preserved and do not directly affect the metric of the geometry. The correspondence between the interior solution in black holes and the cosmological models of an expanding universe give the means to both create and destroy normal and supersymmetric partner particles. This is seen best in the contribution to cosmology of primordial black holes in their formation and evaporation by quantum processes.

 The significant result of this paper is to provide a model for normal and dark matter creation in SUSY and broken SUSY regimes in the early universe , using curved spacetime Bogoliubov transforms of  bosonic and fermionic field operators.[Ref 1 to 22]

\qquad\qquad ACKNOWLEDGEMENTS

I would like to thank Dr Balasubramaniam, the Director of the Institute of Mathematical sciences, Chennai and Prof Sharatchandra for supporting my visit and  I also thank the Institute for its excellent facilities and its members for discussions. 

Almost 34 years ago the author worked on the joint quantisation of Bose and Fermi systems and the manuscript of some 24 pages did not get published, although I gave a long seminar  in november 1972 at IIT , Mumbai. As the encyclopedia of supersymmetry indicates [Ref 2], there were many published and unpublished works in this field. The topic of quantum field theory in curved spacetime also had  several approaches; and my C* algebra approach in a 1975 manuscript  at University of Texas at Austin   was prior to any such  publications by many years. Hence an acknowledgement to those who worked on these subjects.

\qquad\qquad\qquad REFERENCES

1. Birrell N.D., Davies P.C.W. Quantum fields in curved space, Cambridge university press (1984)

2. Duplij Steven, Siegel W., Bagger J. , Encyclopedia of Spersymmetry , Kluwer (2004)
 
3. Varadarajan V.S., Supersymmetry for Mathematicians, Courant lectures v.11, AMS (2004)

4. Freed D. S. , 5 lectures on Supersymmetry AMS (1999)

5. Mishra S.P. , Introduction to Supersymmetry and Supergravity , SERC schools series , Wiley Eastern (1992)

6. Fulling S.A., Aspects of QFT in curved spacetime, London math soc , students series 17 (1989)

7. Peacock John, Cosmological Physics , Cambridge university press (1999)

8. Duck Ian, Sudarshan E.C.G., Pauli and the spin statistics theorem , WSP (1997)

9. Gibbons G., Hawking S., Townsend P. SUSY and its applicationin cosmology , CUP (1996)

10. Khare Avinash, Fractional statistics and quantum theory , 2nd edn , WSP (2005)

11. Khare Avinash, Supersymmetric quantum mechanics , WSP (2001)

12. Iachello F., Interacting Bose Fermi systems in nucclei , Ettore Majorana lectures

    Iachello F., van Isacker P., Ed Cambridge monographs in math phys

13. Keifer Claus, Quantum Gravity,

    International series monographs in physics Oxford science publications (2004)  

14. Berezin F.A. Method of second quantisation, Academic press (1966)

15. Wightman A.S., Streater f, 

              PCT spin statistics and all that, W A Benjamin publ (1968)

16. Freund Peter, Introduction to SUSY, CUP (1989)

17. Patwardhan Ajay, www.arxiv.org/hep-th/0606080 ,

                     www.arxiv.org/hep-th/0406049,

                     www.arxiv.org/quant-ph/0305150,

                     wwww.arxiv.org/hep-th/0611007.

18. Bogoliubov transform papers--

    Elmfors P., Umezawa H.hep-th/9904006,
 
        hep-th/0208183, cond-mat/0109205,

   Gueira H,  daSilva J, Khanna F., Revzen M., Santana A,

    hep-th/0311246,

19. Supersymmetry papers: Pati Jogesh hep-ph/9506211,

                           Fiore Gaetone  hep-th/9611144, 

                          Oeckl R. hep-th/0008072,

20. Supersymmetry and black hole papers:

                            Kallosh Renata, hep-th/9306095, 

                              Green Anne, astro-ph/9903484,
                                        
                                Mazumdar P. , hep-th/9304089,
                                         
                                            hep-th/9904006

21. Supersymmetry and cosmology papers: 

                                Banks T. hep-ph/0408260,

                                 Feng J. , hep-ph/0405215,

22. Bogoliubov transforms and black holes and cosmology papers:

     Farley A.N., D'Eath P.D. , gr-qc/0510043, gr-qc/0510027

                 Sato H., Suzuki H, hep-th/9410092

\end{document}